\documentclass[aps,preprint,prd,nofootinbib]{revtex4}
\usepackage{graphicx}
\usepackage{psfrag}

\usepackage{subfigure}
\usepackage{array}
\usepackage{graphicx}
\usepackage{amsmath}
\usepackage{amssymb}
\usepackage{mathrsfs}
\usepackage{bm}
\usepackage{slashed}
\usepackage{threeparttable}
\usepackage{multirow}
\usepackage[colorlinks, linkcolor=black, anchorcolor=black, citecolor=black]{hyperref}
\usepackage[amsmath,thmmarks,hyperref]{ntheorem}
\usepackage{soul}

\allowdisplaybreaks[4]

\begin{document}

\title{Doubly-charged scalar in rare decays of $B_c$ meson}
\author{Tianhong Wang$^1$\footnote{thwang@hit.edu.cn}, Yue Jiang$^1$\footnote{jiangure@hit.edu.cn}, Zhi-Hui Wang$^2$\footnote{wzh19830606@163.com}, and~Guo-Li Wang$^1$\footnote{gl\_wang@hit.edu.cn}\\}
\address{$^1$Department of Physics, Harbin Institute of Technology, Harbin, 150001, China\\
$^2$School of Electrical $\&$ Information Engineering, Beifang University of Nationalities, Yinchuan, 750021, China}

\baselineskip=20pt

\begin{abstract}

In this paper, we study the lepton number violation processes of $B_c$ meson induced by possible doubly-charged scalars. Both the three-body decay channels and the four-body decay channels are considered. For the former, $Br\times\left(\frac{s_\Delta h_{ij}}{M_\Delta^2}\right)^{-2}$ is of the order of $10^{-7}\sim 10^{-9}$, and for the later channels, $Br\times\left(\frac{s_\Delta h_{ij}}{M_\Delta^2}\right)^{-2}$ is of the order of $10^{-12}\sim 10^{-20}$, where $s_\Delta$, $h_{ij}$, $M_\Delta$ are the constants related to the doubly-charged boson.

\end{abstract}

\maketitle

\section{Introduction}

Doubly-charged Higgs bosons ($\Delta^{\pm\pm}$) have been predicted by the Left-Right symmetric models~\cite{pati, Moha, Sen} as the third component of scalar triplets. If one keeps only this triplet as the new physics beyond the Standard Model (without introducing the right-handed neutrinos), the Type-II see-saw models~\cite{Magg, Laz, Moh81, Cheng} are achieved. This particle is phenomenologically interesting as it can decay to two same-sign charged leptons which indicates the lepton number violation (LNV). It has been searched extensively at the Large Hadron Collider (LHC). Until now, there is no evidence to show their existence, which sets constraints on their masses. For example, the latest result of ATLAS Collaboration shows the lower limit on $m(\Delta^{\pm\pm})$ is $770\sim 870$ GeV for final states with $100\%$ decay to $ee,~e\mu$, and $\mu\mu$~\cite{ATLAS}. And for CMS Collaboration, this lower bound is between $800\sim 820$ GeV~\cite{CMS}.

It is also interesting to investigate the low energy processes with doubly-charged Higgs boson as the intermediate state.  Experimentally, the final particles come from the same vertex because the masses of $W$ and $\Delta^{++}$ bosons are very large. Theoretically, the heavy bosons cannot on the mass shell, their contribution is reflected in effective interaction vertices~\cite{pic97}. Such low energy processes include the rare decays of top quark~\cite{quin13}, $\tau$ lepton~\cite{hays17,quin13,din13}, or charged mesons~\cite{pic97,ma09, bam15}. Surely the branching ratios of these decay modes will be very small due to the large Higgs mass and small coupling constant. However, by comparing the experimental results of the branching ratios of  the LNV processes with the theoretical predictions, one can get the lower bound on the parameters involved in the effective vertices~\cite{quin17}.  One may argue that the Majorana neutrino can also lead to the LNV processes, such as neutrinoless double beta decays in low energy processes. Especially for Majorana neutrinos with masses around GeV scale, as they could be on-shell, the narrow width approximation (NWA) can be applied, which greatly enhances the decay widths of these processes~\cite{atre}. However, it may also be possible that there are only three generations of light Dirac neutrinos in nature. If so, one has to find other mechanisms which could give the same neutrinoless double beta decay signal, and doubly charged Higgs boson will be such a possible alternative. If there are only three generations of light Majorana neutrinos, these LNV processes induced by them are greatly suppressed~\cite{barger, ali} and may have the same order of magnitude as the contribution of the doubly-charged Higgs, which makes the later case be important.

In Ref.~\cite{ma09} and Ref.~\cite{bam15}, the $M_1\rightarrow M_2 l_1^\pm l_2^\pm$ processes induced by the $\Delta^{++}$ with $M_1=B^-,~D^-,~K^-$ are considered. In this work, we will study such processes of $B_c^-$ meson. Moreover, we notice that the LNV four body decay processes of heavy mesons with Majorana neutrinos have been extensively studied in theory~\cite{cas13,yuan13, bao13, dong15, mil16}, while such processes within the doubly-charged Higgs boson formalism have not been investigated yet. So a careful calculation of such channels will be a great supplement for the three-body decay modes. Experimentally, as LHCb will produce more and more $B_c$ mesons, searching such decay channels will setting an experimental upper limit for the branching ratios, which can also be used to constrain the parameters of doblely-charged Higgs boson.

This work is organized as follows. In Section II, we present the theoretical formalism. Three-body decay processes and four-body decay processes are both considered. In Section III, we give the numerical results and discussions. Finally, the conclusion is given in Section IV. And some details for the calculation of the hadronic transition matrix element is presented in the Appendix.

\section{Theoretical Formalism}

The Lagrangian describing the interaction of doubly-charged scalars with Standard Model fermions has the form~\cite{pic97}
\begin{equation}
\begin{aligned}
\mathcal L_{int} = ih_{ij}\psi_{iL}^TC\sigma_2\Delta\psi_{jL} + H.c.,
\end{aligned}
\end{equation}
where $\psi_{iL}$ is the two-component leptonic doublet; $h_{ij}$ is the leptonic Yukawa coupling constant; $C=i\gamma^2\gamma^0$ is the charge conjugation matrix; $\sigma_2$ is the second Pauli matrix; $\Delta$ is the complex triplet in the $2\times 2$ representation which we have defined as
\begin{equation}
\Delta = \left(
\begin{array}{cc}
\Delta^+/\sqrt{2}& \Delta^{++} \\
\Delta^0 & -\Delta^+/\sqrt{2}\\
\end{array}
\right).
\end{equation}
The Lagrangian which describes the interaction of $\Delta^{++}/\Delta^\pm$ with $W^-$ gauge boson and quarks has the following form~\cite{pic97, ma09}
\begin{equation}
\begin{aligned}
\mathcal L_{int}^\prime&=-\sqrt{2}gm_Ws_\Delta \Delta^{++}W^{-\mu}W^-_\mu+ \frac{\sqrt{2}}{2}gc_\Delta W^{-\mu}\Delta^-\overset{\leftrightarrow}{\partial}_\mu\Delta^{++}\\
&~~~~+\frac{igs_\Delta}{\sqrt{2}m_Wc_\Delta}\Delta^+(m_{q^\prime}\bar q_Rq_R^\prime-m_q\bar q_Lq_L^\prime)+ H.c.,
\end{aligned}
\end{equation}
where $s_\Delta=\sin\theta_\Delta$ and $c_\Delta=\cos\theta_\Delta$ with $\theta_\Delta$ is the mixing angle between the usual ${\rm SU(2)}_L$ Higgs doublet and the assumed Higgs triplet. 

\subsection{The $B_c^-\rightarrow h^+l_1^-l_2^-$ processes}

The three-body decay process of $B_c^-$ with lepton number violation is shown in Fig.~1. Actually, there are six other diagrams which contain $\Delta^\pm$. However, the contribution of those diagrams is very small compared with those of Fig.~1. This can be seen from that the parameters of the last two terms in Eq.~(3) are very small compared with that of the first term. In Ref.~\cite{quin13}, the ratio of the amplitudes with and without $\Delta^\pm$ is estimated to be less than $10^{-7}$. So here we can safely neglect their contribution. The amplitude corresponding to the two diagrams in Fig.~1 is

\begin{figure}[ht]
\centering
\renewcommand{\thesubfigure}{(\Alph{subfigure})}
\subfigure[]{\includegraphics[scale=0.39]{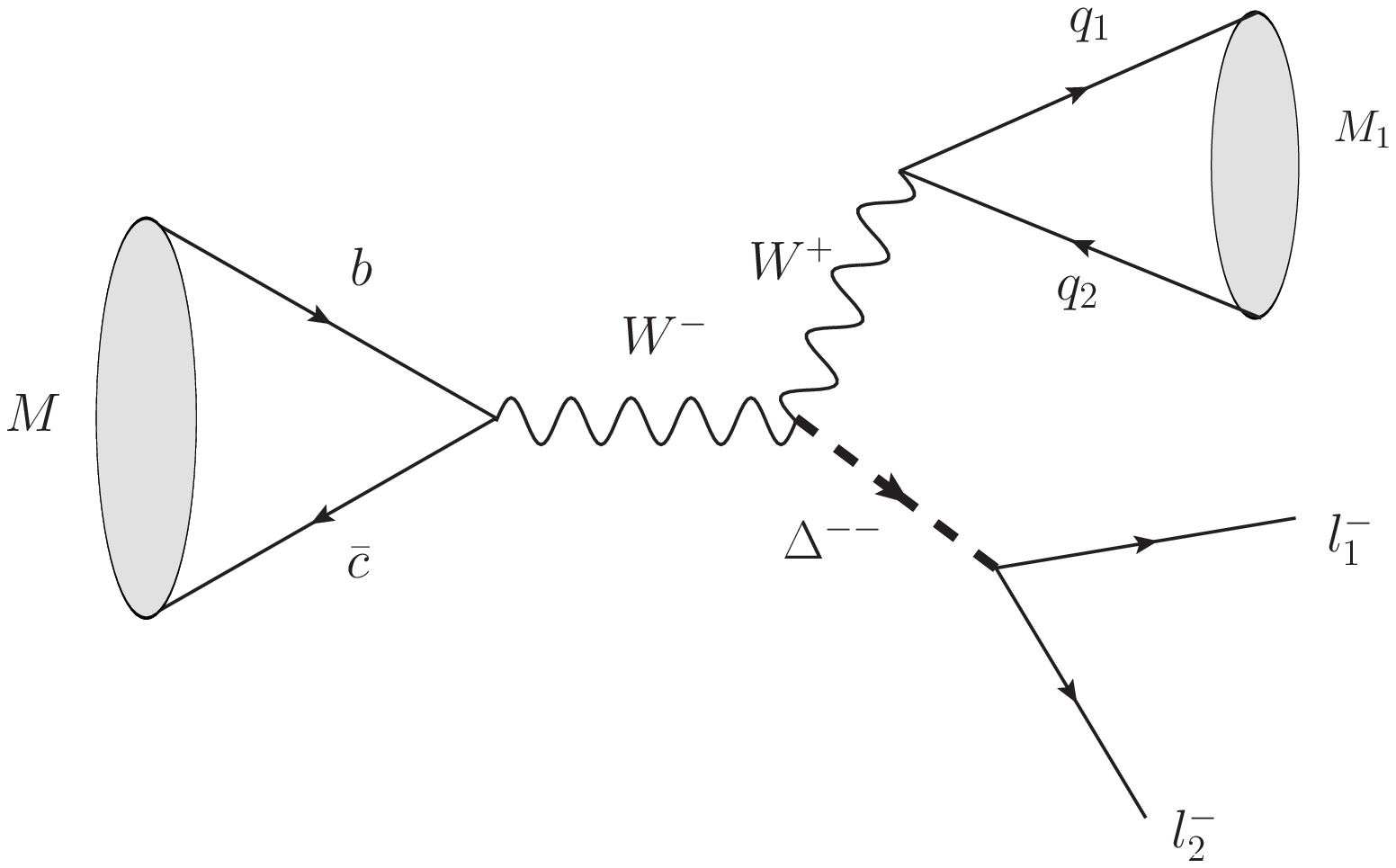}}
\hspace{2cm}
\subfigure[]{\includegraphics[scale=0.39]{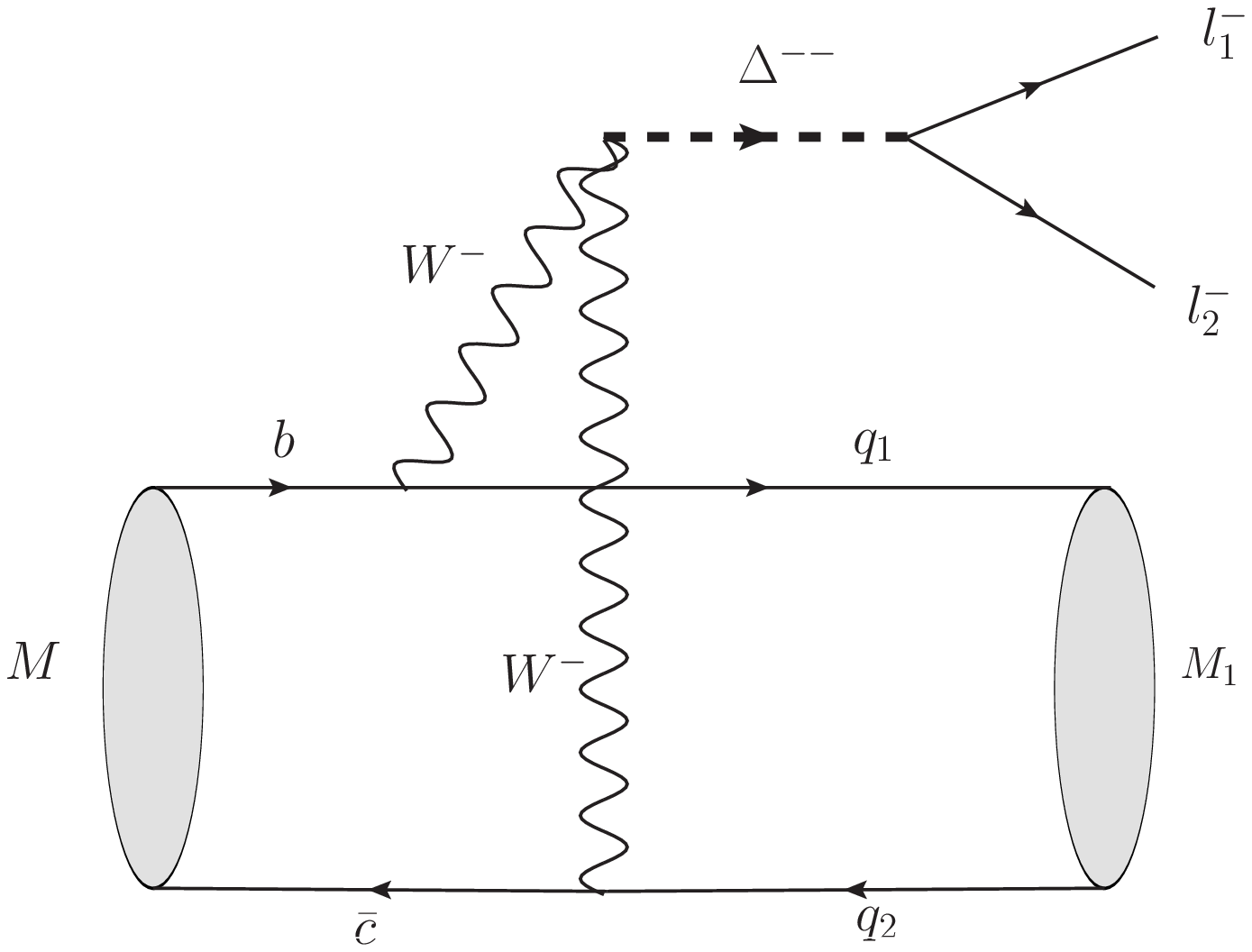}}
\caption[]{Feynman diagrams of the decay processes $B_c^-\rightarrow h^+l_1^-l_2^-$.}
\end{figure}

\begin{equation}
\begin{aligned}
\mathcal M&=\frac{g^3s_\Delta h_{ij}}{8\sqrt{2}m_W^3m_{\Delta}^2}(V_{cb}^\ast V_{q_1q_2}+\frac{1}{3}V_{q_1b}V_{cq_2}^\ast)\langle h(p_1)| (\bar cb)_{_{V-A}} (\bar q_1q_2)_{_{V-A}} |B_c^-(p)\langle lepton\rangle\\
&=\frac{g^3s_\Delta h_{ij}}{8\sqrt{2}m_W^3m_{\Delta}^2}(V_{cb}^\ast V_{q_1q_2}+\frac{1}{3}V_{q_1b}V_{cq_2}^\ast)f_h f_{B_c} p\cdot p_1\langle lepton\rangle,
\end{aligned}
\end{equation}
where we have used the definition $\langle h(p_1)| \bar q_1\gamma^\mu(1-\gamma_5) q_2|0\rangle = if_h p_1^\mu$ with $f_h$ being the decay constant of the final pseudoscalar meson. For the vector meson case, the definition $\langle h(p_1, \epsilon)| \bar q_1\gamma^\mu(1-\gamma_5) q_2|0\rangle = f_h M_1\epsilon^\mu_1$ should be applied, and in Eq.~(4), $p\cdot p_1$ should be changed to $M_1p\cdot \epsilon_1$. We also defined $\langle lepton\rangle \equiv \bar v(k_2)(1-\gamma_5)u(k_1)-\bar v(k_1)(1-\gamma_5)u(k_2)$, where $u(k_i)$ and $v(k_i)$ are the spinors of charged leptons. The factor $\frac{1}{3}$ in the parentheses comes from the Fierz transformation.
The squared amplitude can be written as
\begin{equation}
\begin{aligned}
|\mathcal M|^2&=\sqrt{2}G_F^3\left(\frac{s_\Delta h_{ij}}{M_\Delta^2}\right)^2|V_{cb}^\ast V_{q_1q_2}+\frac{1}{3}V_{q_1b}V_{cq_2}^\ast|^2f_h^2 f_{B_c}^2 |p\cdot p_1\langle lepton\rangle|^2,
\end{aligned}
\end{equation}
where we have used the definition $\frac{G_F}{\sqrt{2}}=\frac{g^2}{8m_W^2}$.

The partial decay width can be achieved by finishing the phase space integral
\begin{equation}
\Gamma = (1-\frac{1}{2}\delta_{l_1l_2})\frac{1}{512\pi^3M^3}\int\frac{ds_{12}}{s_{12}}\lambda^{1/2}(M^2, s_{12}, M_1^2)\lambda^{1/2}(s_{12}, m_1^2, m_2^2)\int d\cos\theta_{12}{|\mathcal M|^2},
\end{equation}
where $s_{12}\equiv(k_1+k_2)^2$; $m_1$ and $m_2$ are the masses of two charged lepton $l_1$ and $l_2$, respectively; the K${\rm \ddot a}$llen function 
\begin{equation}
\lambda(x, y, z)= x^2 + y^2 +z^2 -2xy-2xz -2yz
\end{equation}
is used; $\theta_{12}$ is the angle between the three-momenta $\vec k_{12}= \vec k_1 + \vec k_2$ and $\vec K_1$ (the later is the three-momentum of $l_1$ in the center-of-momentum frame of $l_1$ and $l_2$). $\delta_{l_1l_2}=0~(1)$ when $l_1$ and $l_2$ are nonidentical (identical) leptons. The integral limits are 
\begin{equation}
\begin{aligned}
s_{12}\in[(m_1+m_2)^2,~(M-M_1)^2],~~~~~\theta_{12}\in[0,~\pi].
\end{aligned}
\end{equation}

\subsection{The $B_c^-\rightarrow h_1^0h_2^+l_1^-l_2^-$ processes}

\begin{figure}[ht]
\centering
\renewcommand{\thesubfigure}{(\Alph{subfigure})}
\subfigure[]{\includegraphics[scale=0.39]{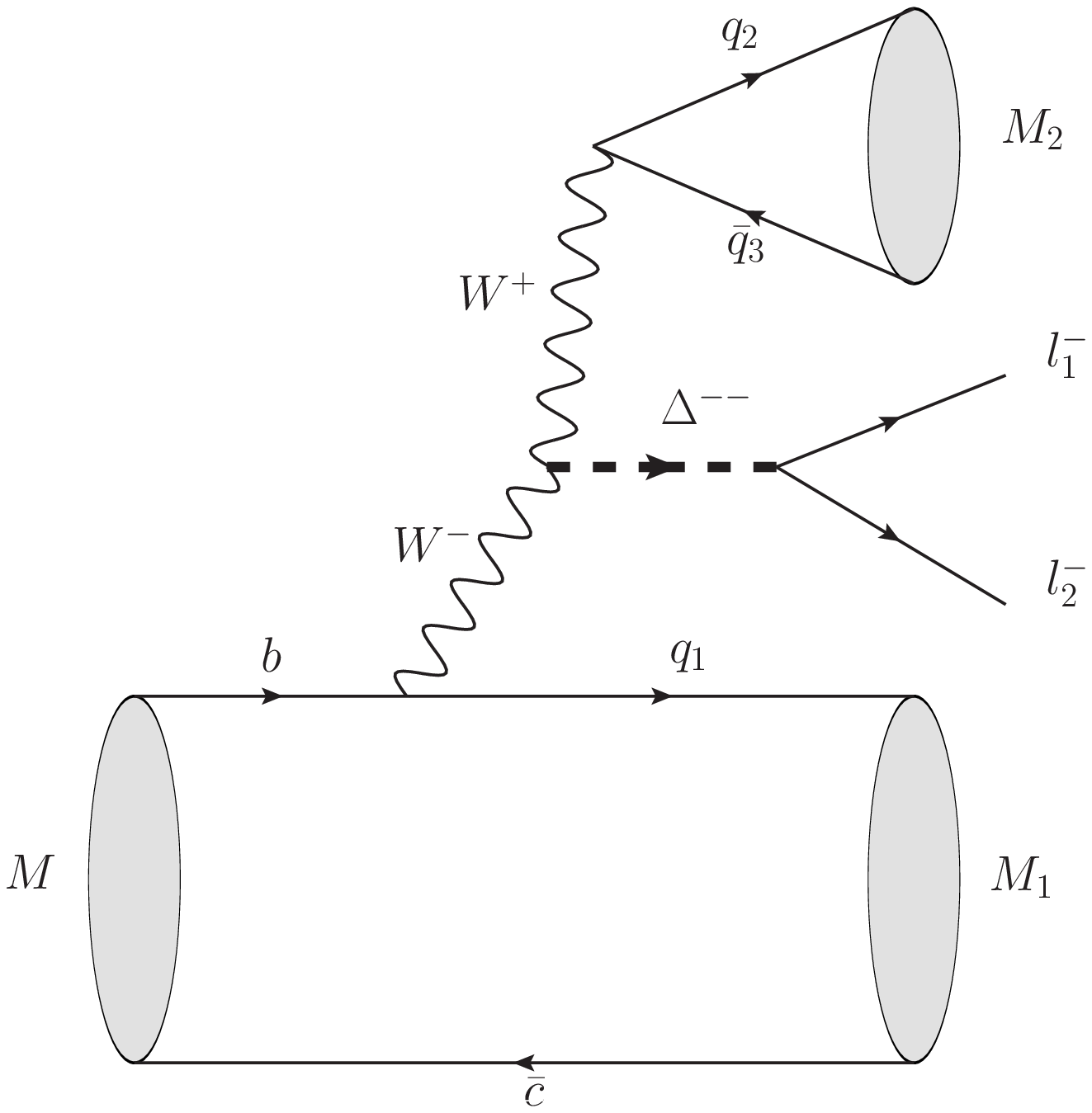}}
\hspace{2cm}
\subfigure[]{\includegraphics[scale=0.39]{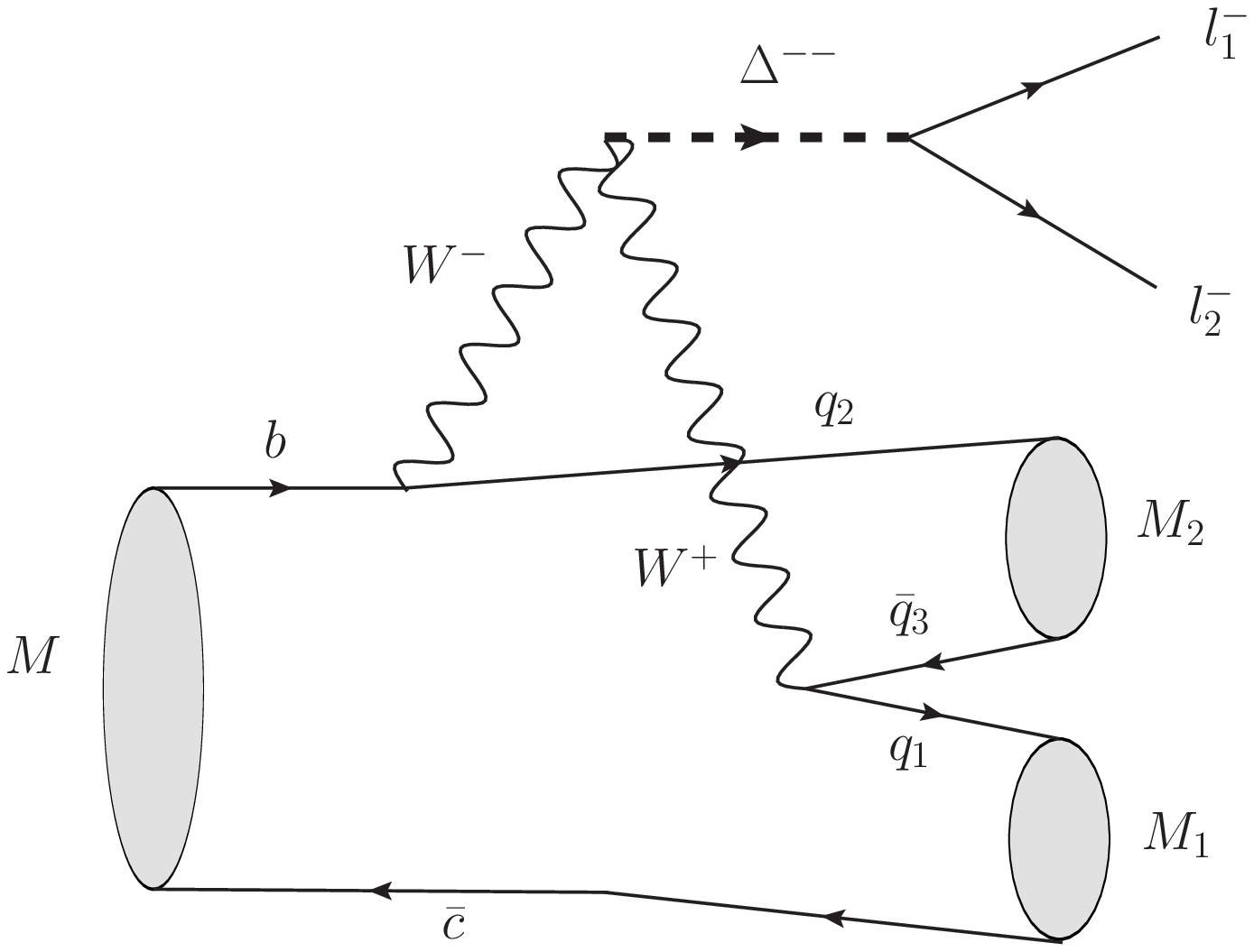}}
\subfigure[]{\includegraphics[scale=0.39]{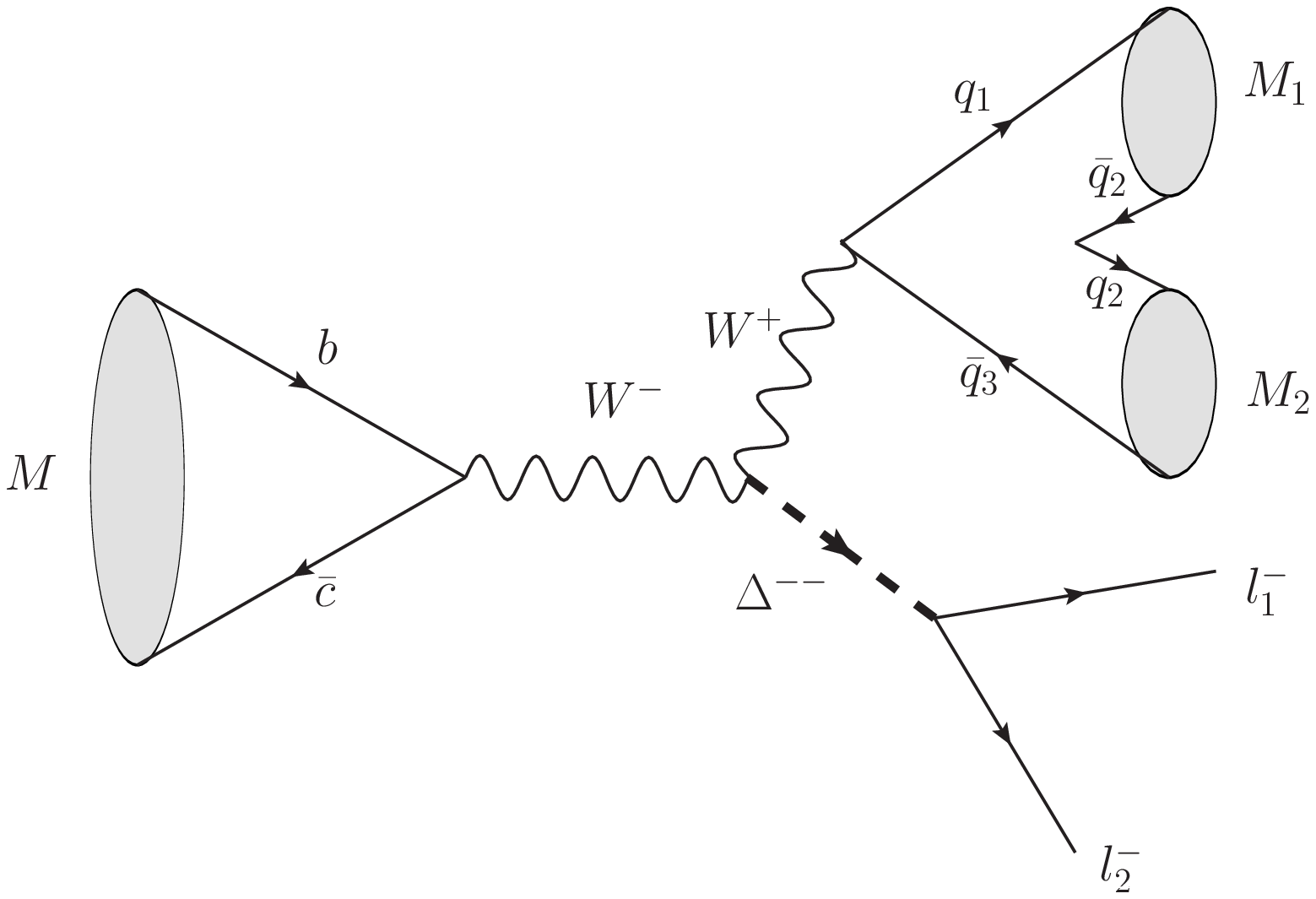}}
\hspace{2cm}
\subfigure[]{\includegraphics[scale=0.39]{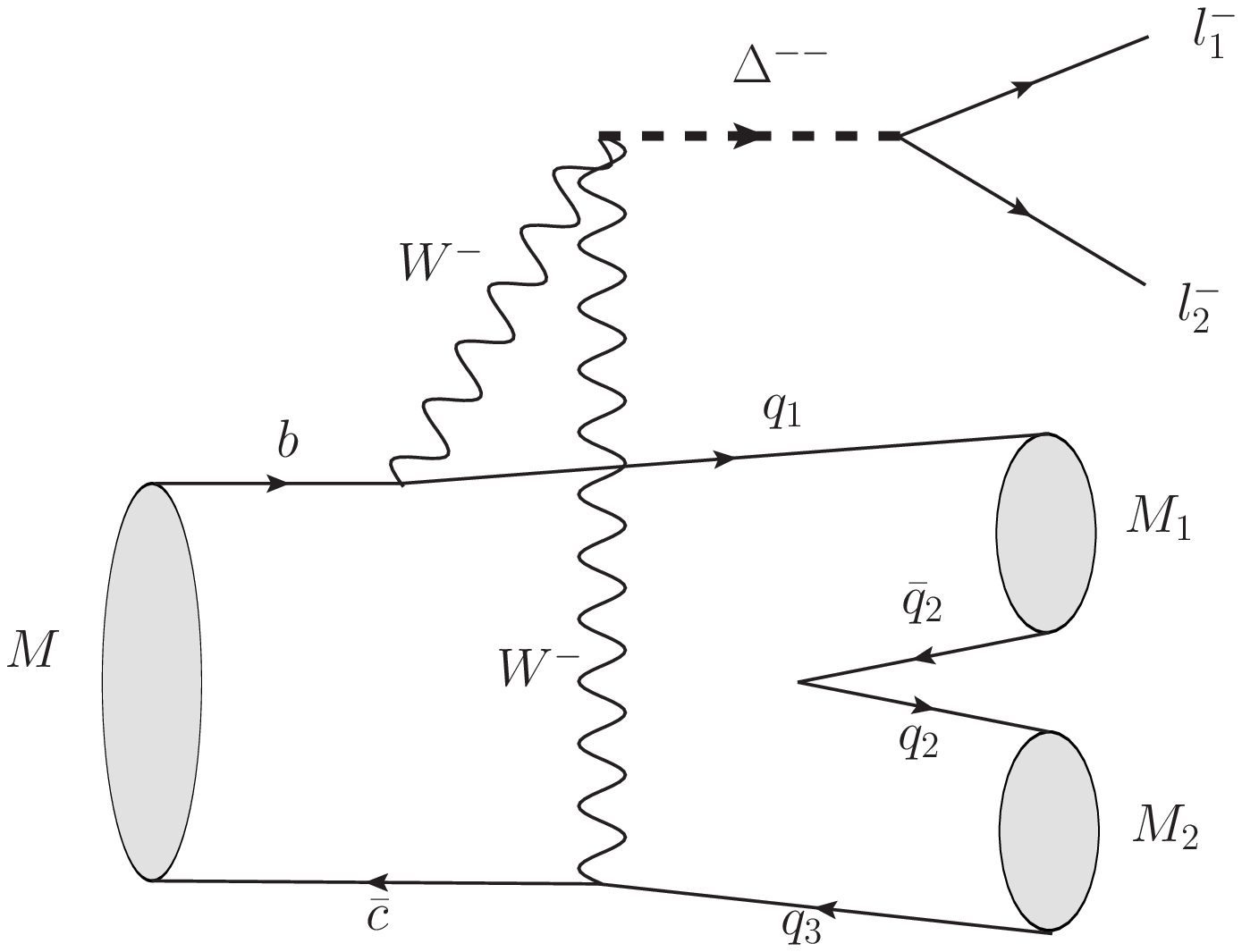}}
\caption[]{Feynman diagrams of the decay processes $B_c^-\rightarrow h_1^0h_2^+l_1^-l_2^-$}
\end{figure}

 For the $B_c^-\rightarrow J/\psi h_2^+l_1^-l_2^-$ processes, when $h_2^+=\pi^+$ or $K^+$, only the Feynman diagrams of Figs. 2(A) and 2(B) contribute; when $h_2^+=D^+,~D_s^+$, all the four diagrams of Fig. 2 give contribution, while that of (C) and (D) could be neglected as the $c\bar c$ pair production will be highly suppressed. So we only consider the contribution of (A) and (B). The corresponding amplitudes are written as
\begin{equation}
\begin{aligned}
\mathcal M_{A} &= \frac{g^3}{8\sqrt{2}m_W^3}V_{cb}V_{q_2q_3} \frac{s_\Delta h_{ij}}{m_{\Delta}^2}\langle J/\psi(p_1)h_2(p_2)| (\bar cb)_{_{V-A}} (\bar q_2q_3)_{_{V-A}} |B_c^-(p)\rangle \langle lepton\rangle\\
&=\frac{g^3}{8\sqrt{2}m_W^3}V_{cb}V_{q_2q_3} \frac{s_\Delta h_{ij}}{m_{\Delta}^2}f_{h_2} p_2^\mu\langle J/\psi(p_1)| \bar c\gamma_\mu(1-\gamma_5)b |B_c^-(p)\rangle \langle lepton\rangle,
\end{aligned}
\end{equation}
\begin{equation}
\begin{aligned}
\mathcal M_{B} &=\frac{g^3}{8\sqrt{2}m_W^3}V_{q_2b}V_{cq_3} \frac{s_\Delta h_{ij}}{m_{\Delta}^2}\langle J/\psi(p_1)h_2(p_2)| (\bar q_2b)_{_{V-A}} (\bar cq_3)_{_{V-A}} |B_c^-(p)\rangle \langle lepton\rangle\\
&=\frac{g^3}{24\sqrt{2}m_W^3}V_{q_2b}V_{cq_3} \frac{s_\Delta h_{ij}}{m_{\Delta}^2}f_{h_2} p_2^\mu\langle J/\psi(p_1)| \bar c\gamma_\mu(1-\gamma_5)b |B_c(p)\rangle \langle lepton\rangle,
\end{aligned}
\end{equation}
where we have used the Fierz transformation in $\mathcal M_B$. Here we only give the results when $h_2$ is a pseudoscalar meson. If $h_2$ is a vector meson, $f_{h_2} p_2^\mu$ should be replaced by $M_2f_{h_2}\epsilon_2^\mu$  Finally, we get the transition amplitude
\begin{equation}
\begin{aligned}
\mathcal M&=\frac{g^3s_\Delta h_{ij}}{8\sqrt{2}m_W^3m_{\Delta}^2}(V_{cb}V_{q_2q_3}+\frac{1}{3}V_{q_2b}V_{cq_3})f_{h_2} p_2^\mu\langle J/\psi(p_1)| \bar c\gamma_\mu(1-\gamma_5)b |B_c^-(p)\rangle\langle lepton\rangle.
\end{aligned}
\end{equation}
The hadronic transition matrix can be expressed as~\cite{fu12}
\begin{equation}
\begin{aligned}
&\langle J/\psi(p_1)| V^\mu |B_c^-(p)\rangle= -i\frac{2}{M+M_1}f_V(Q^2)\epsilon^{\mu\epsilon^\ast pp_1},\\
&\langle J/\psi(p_1)| A^\mu |B_c^-(p)\rangle= f_1(Q^2)\frac{\epsilon^\ast\cdot p}{M+M_1}(p+p_1)^\mu+f_2(Q^2)\frac{\epsilon^\ast\cdot p}{M+M_1}(p-p_1)^\mu\\
&~~~~~~~~~~~~~~~~~~~~~~~~~~~~~~+ f_0(Q^2)(M+M_1)\epsilon^{\ast\mu},
\end{aligned}
\end{equation}
where $Q=p-p_1$, $f_i$ ($i=0,~1,~2$) are form factors.

For the $B_c^-\rightarrow \bar D^{(\ast)0} h_2^+l_1^-l_2^-$ processes, $h_2^+$ can also be $\pi^+$, $K^+$, $D^+$, or $D_s^+$. For the same reason as the $J/\psi$ case, the contribution of Fig.~2(C) and (D) for $D^+$ and $D_s^+$ situations are also neglected. The transition amplitude can be written as
\begin{equation}
\begin{aligned}
\mathcal M&=\frac{g^3s_\Delta h_{ij}}{8\sqrt{2}m_W^3m_{\Delta}^2}(V_{ub}V_{q_2q_3}+\frac{1}{3}V_{q_2b}V_{uq_3})f_{h_2} p_2^\mu\langle \bar D^{(\ast)0}(p_1)| \bar c\gamma_\mu(1-\gamma_5)b |B_c^-(p)\rangle\langle lepton\rangle.
\end{aligned}
\end{equation}
For $\bar D^{\ast0}$, the hadronic transition matrix element is parameterized in the same way as Eq.~(12). For $\bar D^0$, it is parameterized as~\cite{fu12}
\begin{equation}
\begin{aligned}
\langle \bar D^0(p_1)| V^\mu |B_c^-(p)\rangle&= f_+(Q^2)(p+p_1)^\mu+f_-(Q^2)(p-p_1)^\mu,
\end{aligned}
\end{equation}
where $f_\pm$ are form factors.

The phase space integral for four body decay processes can be expressed as 
\begin{equation}
\begin{aligned}
\Gamma = (1-\frac{1}{2}\delta_{l_1l_2})\int \frac{ds_{12}}{s_{12}}\int\frac{ds_{34}}{s_{34}} \int d\cos\theta_{12} \int d\cos\theta_{34} \int d\phi  \mathcal K|\mathcal M|^2,
\end{aligned}
\end{equation}
where $s_{12}=(p_1+p_2)^2$, $s_{34}=(p_3+p_4)^2$. $\theta_{12}$ is the angle between the three-momenta $\vec p_{12}=\vec p_1 +\vec p_2$  and $\vec P_1$ (the later is the three-momentum of $h_1$ in the center-of-momentum frame of $h_1$ and $h_2$). $\theta_{34}$ is the angle between the three-momenta $\vec k_{12}=\vec k_1+\vec k_2$ and $\vec K_1$ (the later is the three-momentum of $l_1$ in the center-of-momentum frame of $l_1$ and $l_2$). $\phi$ is the angle between the decay planes $\Sigma(h_1h_2)$ and $\Sigma(l_1l_2)$. The factor $\mathcal K$ has the expression
\begin{equation}
\mathcal K = \frac{1}{2^{15}\pi^6 M^3} \lambda^{1/2}(M^2, s_{12}, s_{34}) \lambda^{1/2}(s_{12}, M_1^2, M_2^2)\lambda^{1/2}(s_{34},m_1^2, m_2^2) .
\end{equation}
The integral limits are
\begin{equation}
\begin{aligned}
&s_{12}\in[(M_1+M_2)^2,~(M-m_1-m_2)^2],\\
&s_{34}\in[(m_1+m_2)^2,~(M-\sqrt{s_{12}})^2],\\
&\phi\in[0,~2\pi],~~~\theta_{12}\in[0,~\pi],~~~\theta_{34}\in[0,~\pi].
\end{aligned}
\end{equation}

\section{Numerical Results}

Here we present some parameters used in the calculation. The lifetime of $B_c^-$ meson is $0.507\times 10^{-12}$ s~\cite{pdg}. The decay constants used here are as follows: $f_{B_c}=0.322$ GeV~\cite{cve04}, $f_\pi=130.4$ MeV, $f_K=156.2$ MeV, $f_D=204.6$ MeV, and $f_{D_s}=257.5$ MeV~\cite{pdg}, $f_\rho=0.205$ GeV, $f_{K^\ast}=0.217$ GeV~\cite{ball}, $f_{D^\ast}=0.340$ GeV, and $f_{D_s^\ast}=0.375$ GeV~\cite{wang06}. The quark masses used here are: $m_b=4.96$ GeV, $m_c=1.62$ GeV, $m_s=1.50$ GeV, $m_d=0.311$ GeV, and $m_u=0.305$ GeV.

The branching ratios of three body decay channels is presented in Table I. Here we have divided the factor $\left(\frac{s_\Delta h_{ij}}{M_\Delta^2}\right)^{2}$. One can see the $D_s^+l^-l^-$ channel has the largest value which is of the order of $10^{-7}$. The channel with $l^-=\mu^-$ has almost the same width as that of $l^-=e^-$, which means the process is insensitive to the lepton mass. The channel with $\mu^- e^-$ as the final leptons has width about 2 times of that of the former two channels because of $\delta_{l_1l_2}=0$ for this case. To estimate the upper limit of the decay width, we have to give the lower limit of mass of the doubly charged Higgs boson and the upper limit of the coupling constant $h_{ij}$. If we take the same values in Ref.~\cite{ma09}, that is 
\begin{equation}
\begin{aligned}
&h_{ee}^2=9.7 \times 10^{-6}{\rm GeV}^{-2}M_{\Delta}^2,\\
&h_{\mu\mu}^2=2.5 \times 10^{-5}{\rm GeV}^{-2}M_{\Delta}^2,\\
&h_{\mu e}^2=1.6 \times 10^{-15}{\rm GeV}^{-2}M_{\Delta}^2,\\
&s_\Delta<0.0056,\\
\end{aligned}
\end{equation}
and set $M_{\Delta}\sim 1000$ GeV considering the latest results in Refs.~\cite{CMS, ATLAS}, then we can get $\left(\frac{s_\Delta h_{ij}}{M_{\Delta}^2}\right)^{2}<3.0\times 10^{-16}$  ${\rm GeV}^{-4}$ (for $ee$), $7.8\times10^{-16}$ ${\rm GeV}^{-4}$ (for $\mu\mu$), $5.0\times10^{-26}$ ${\rm GeV}^{-4}$ (for $\mu e$). So the largest upper limit of the branching ratios is of the order of $10^{-23}$.

In Tabel II, we give the branching ratios of four body decay channels with $J/\psi$ as one of the final mesons. Compared with the three-body decay channels, the branching ratios here are several orders smaller. Actually, most of the suppression comes from the phase space integral. We can estimate this as follows: from Eq.~(6) and Eq.~(15) one can see that the ratio of the constants is $(2^6\pi^3)^{-1}=5.0\times10^{-4}$, which provide most of the difference between Table I and II. The channels which have the largest upper limit are $J/\psi D_s^{\ast+}l^-l^-$, which are about $10^{-29}$ (by using the values in Eq.~(18)). One noticed that, in Refs.~\cite{mil16, mad16},  $B_c$ four-body decays with a GeV scale Majorana neutrino are calculated. There Fig. 2(B) gives negligible contribution. Here this diagram is just color suppressed, while its contribution can have the same order of magnitude as that of Fig. 2(A). 

For the $B_c^-\rightarrow \bar D^{(\ast)0} h_2^+l_1^-l_2^-$ channels, the results are given in Table III and Table IV. The largest upper limit of decay widths for these channels is of the order of $10^{-29}$. As the final states contain $\bar c$, only Fig.~2(A) and (B) contribute to the channels with $h_2^+=\pi^+,~K^+$. For the channels with $h_2^+=D^+,~D_s^+$, (C) and (D) also give contribution, while they are neglected for the reason above. One notices that the decay widths in Table IV are about one order less than those in Table III. This is different with the semi-leptonic decay channels of $B$~\cite{pdg}, where the $D^{\ast0}l^-\nu_l$ channel has larger width than that of $D^{0}l^-\nu_l$. 
In Table V and Table VI, we present the branching ratios of channels with $\bar B_{(s)}^0h_1^+l_1^-l_2^-$ and $\bar B_{(s)}^{\ast 0}h_1^+l_1^-l_2^-$ as the final states. The decay width of the $\bar B_s^0\pi^+l^-l^-$ channel has the largest upper limit of $10^{-28}$, which mainly due to the large CKM matrix elements.

Here three things should be mentioned to the four-body decay channels. First, except the channels calculated here, there are also some other channels which can only be realized through Fig. 2 (C) and (D), such as $D^{(\ast)0}h_2^+l_1^-l_2^-$ and $\pi^+\pi^0l_1^-l_2^-$ channels. They are not considered here. Second, the QCD corrections are not considered here. But it is easy to be added if only Fig.~(A) and (B) contribute, which is similar to that of the two body nonleptonic decay channels of the $B_c$ meson. Third, the final state interaction (FSI) are not considered here, since it will not greatly change the results' order of magnitude.

\begin{table}
\caption{Branching ratios of three-body decay channels of $B_c^-$.}
\vspace{0.2cm}
\setlength{\tabcolsep}{0.01cm}
\centering
\begin{tabular*}{\textwidth}{@{}@{\extracolsep{\fill}}cccc}
\hline\hline
decay channel& $Br\times\left(\frac{s_\Delta h_{ij}}{M_\Delta^2}\right)^{-2}$&decay channel& $Br\times\left(\frac{s_\Delta h_{ij}}{M_\Delta^2}\right)^{-2}$\\ \hline
{\phantom{\Large{l}}}\raisebox{+.2cm}{\phantom{\Large{j}}}
$B_c^-\rightarrow \pi^+e^-e^-$&$2.47\times 10^{-8}$&$B_c^-\rightarrow \rho^+e^-e^-$&$4.98\times 10^{-8}$\\
{\phantom{\Large{l}}}\raisebox{+.2cm}{\phantom{\Large{j}}}
$B_c^-\rightarrow \pi^+\mu^-\mu^-$&$2.45\times 10^{-8}$&$B_c^-\rightarrow \rho^+\mu^-\mu^-$&$4.95\times 10^{-8}$\\
{\phantom{\Large{l}}}\raisebox{+.2cm}{\phantom{\Large{j}}}
$B_c^-\rightarrow \pi^+e^-\mu^-$&$4.92\times 10^{-8}$&$B_c^-\rightarrow \rho^+e^-\mu^-$&$9.94\times 10^{-8}$\\
{\phantom{\Large{l}}}\raisebox{+.2cm}{\phantom{\Large{j}}}
$B_c^-\rightarrow K^+e^-e^-$&$1.95\times 10^{-9}$&$B_c^-\rightarrow K^{\ast+}e^-e^-$&$2.92\times 10^{-9}$\\
{\phantom{\Large{l}}}\raisebox{+.2cm}{\phantom{\Large{j}}}
$B_c^-\rightarrow K^+\mu^-\mu^-$&$1.94\times 10^{-9}$&$B_c^-\rightarrow K^{\ast+}\mu^-\mu^-$&$2.90\times 10^{-9}$\\
{\phantom{\Large{l}}}\raisebox{+.2cm}{\phantom{\Large{j}}}
$B_c^-\rightarrow  K^+e^-\mu^-$&$3.89\times 10^{-9}$&$B_c^-\rightarrow  K^{\ast+}e^-\mu^-$&$5.82\times 10^{-9}$\\
{\phantom{\Large{l}}}\raisebox{+.2cm}{\phantom{\Large{j}}}
$B_c^-\rightarrow D^+e^-e^-$&$3.98\times 10^{-9}$&$B_c^-\rightarrow D^{\ast+}e^-e^-$&$4.45\times 10^{-9}$\\
{\phantom{\Large{l}}}\raisebox{+.2cm}{\phantom{\Large{j}}}
$B_c^-\rightarrow  D^+\mu^-\mu^-$&$3.95\times 10^{-9}$&$B_c^-\rightarrow  D^{\ast+}\mu^-\mu^-$&$4.41\times 10^{-9}$\\
{\phantom{\Large{l}}}\raisebox{+.2cm}{\phantom{\Large{j}}}
$B_c^-\rightarrow D^+e^-\mu^-$&$7.94\times 10^{-9}$&$B_c^-\rightarrow D^{\ast+}e^-\mu^-$&$8.85\times 10^{-9}$\\
{\phantom{\Large{l}}}\raisebox{+.2cm}{\phantom{\Large{j}}}
$B_c^-\rightarrow D_s^+e^-e^-$&$1.18\times 10^{-7}$&$B_c^-\rightarrow D_s^{\ast+}e^-e^-$&$9.39\times 10^{-8}$\\
{\phantom{\Large{l}}}\raisebox{+.2cm}{\phantom{\Large{j}}}
$B_c^-\rightarrow D_s^+\mu^-\mu^-$&$1.17\times 10^{-7}$&$B_c^-\rightarrow D_s^{\ast+}\mu^-\mu^-$&$9.31\times 10^{-8}$\\
{\phantom{\Large{l}}}\raisebox{+.2cm}{\phantom{\Large{j}}}
$B_c^-\rightarrow D_s^+e^-\mu^-$&$2.35\times 10^{-7}$&$B_c^-\rightarrow D_s^{\ast+}e^-\mu^-$&$1.87\times 10^{-7}$\\
\hline\hline
\end{tabular*}
\end{table}

\begin{table}
\caption{Branching ratios of $B_c^-$ decays induced by the current $(\bar cb)_{_{V-A}}(\bar q_1q_2)_{_{V-A}}$.}
\vspace{0.2cm}
\setlength{\tabcolsep}{0.01cm}
\centering
\begin{tabular*}{\textwidth}{@{}@{\extracolsep{\fill}}cccc}
\hline\hline
decay channel&$Br\times\left(\frac{s_\Delta h_{ij}}{M_\Delta^2}\right)^{-2}$&decay channel& $Br\times\left(\frac{s_\Delta h_{ij}}{M_\Delta^2}\right)^{-2}$\\ \hline
{\phantom{\Large{l}}}\raisebox{+.2cm}{\phantom{\Large{j}}}
$B_c^-\rightarrow J/\psi \pi^+e^-e^-$&$2.24\times 10^{-14}$&$B_c^-\rightarrow J/\psi \rho^+e^-e^-$&$9.25\times 10^{-14}$\\
{\phantom{\Large{l}}}\raisebox{+.2cm}{\phantom{\Large{j}}}
$B_c^-\rightarrow J/\psi \pi^+\mu^-\mu^-$&$2.37\times 10^{-14}$&$B_c^-\rightarrow J/\psi \rho^+\mu^-\mu^-$&$1.01\times 10^{-13}$\\
{\phantom{\Large{l}}}\raisebox{+.2cm}{\phantom{\Large{j}}}
$B_c^-\rightarrow J/\psi \pi^+e^-\mu^-$&$4.58\times 10^{-14}$&$B_c^-\rightarrow J/\psi \rho^+e^-\mu^-$&$1.93\times 10^{-13}$\\
{\phantom{\Large{l}}}\raisebox{+.2cm}{\phantom{\Large{j}}}
$B_c^-\rightarrow J/\psi K^+e^-e^-$&$3.34\times 10^{-15}$&$B_c^-\rightarrow J/\psi K^{\ast+}e^-e^-$&$6.96\times 10^{-15}$\\
{\phantom{\Large{l}}}\raisebox{+.2cm}{\phantom{\Large{j}}}
$B_c^-\rightarrow J/\psi K^+\mu^-\mu^-$&$3.60\times 10^{-15}$&$B_c^-\rightarrow J/\psi K^{\ast+}\mu^-\mu^-$&$7.63\times 10^{-15}$\\
{\phantom{\Large{l}}}\raisebox{+.2cm}{\phantom{\Large{j}}}
$B_c^-\rightarrow J/\psi K^+e^-\mu^-$&$6.89\times 10^{-15}$&$B_c^-\rightarrow J/\psi K^{\ast+}e^-\mu^-$&$1.45\times 10^{-14}$\\
{\phantom{\Large{l}}}\raisebox{+.2cm}{\phantom{\Large{j}}}
$B_c^-\rightarrow J/\psi D^+e^-e^-$&$9.66\times 10^{-15}$&$B_c^-\rightarrow J/\psi D^{\ast+}e^-e^-$&$1.94\times 10^{-14}$\\
{\phantom{\Large{l}}}\raisebox{+.2cm}{\phantom{\Large{j}}}
$B_c^-\rightarrow J/\psi D^+\mu^-\mu^-$&$1.07\times 10^{-14}$&$B_c^-\rightarrow J/\psi D^{\ast+}\mu^-\mu^-$&$2.17\times 10^{-14}$\\
{\phantom{\Large{l}}}\raisebox{+.2cm}{\phantom{\Large{j}}}
$B_c^-\rightarrow J/\psi D^+e^-\mu^-$&$2.00\times 10^{-14}$&$B_c^-\rightarrow J/\psi D^{\ast+}e^-\mu^-$&$4.02\times 10^{-14}$\\
{\phantom{\Large{l}}}\raisebox{+.2cm}{\phantom{\Large{j}}}
$B_c^-\rightarrow J/\psi D_s^+e^-e^-$&$2.69\times 10^{-13}$&$B_c^-\rightarrow J/\psi D_s^{\ast+}e^-e^-$&$3.99\times 10^{-13}$\\
{\phantom{\Large{l}}}\raisebox{+.2cm}{\phantom{\Large{j}}}
$B_c^-\rightarrow J/\psi D_s^+\mu^-\mu^-$&$2.98\times 10^{-13}$&$B_c^-\rightarrow J/\psi D_s^{\ast+}\mu^-\mu^-$&$4.44\times 10^{-13}$\\
{\phantom{\Large{l}}}\raisebox{+.2cm}{\phantom{\Large{j}}}
$B_c^-\rightarrow J/\psi D_s^+e^-\mu^-$&$5.55\times 10^{-13}$&$B_c^-\rightarrow J/\psi D_s^{\ast+}e^-\mu^-$&$8.21\times 10^{-13}$\\
\hline\hline
\end{tabular*}
\end{table}

\begin{table}
\caption{Branching ratios of $B_c^-$ decays induced by the current $(\bar ub)_{_{V-A}}(\bar q_1q_2)_{_{V-A}}$.}
\vspace{0.2cm}
\setlength{\tabcolsep}{0.01cm}
\centering
\begin{tabular*}{\textwidth}{@{}@{\extracolsep{\fill}}cccc}
\hline\hline
decay channel&$Br\times\left(\frac{s_\Delta h_{ij}}{M_\Delta^2}\right)^{-2}$&decay channel&$Br\times\left(\frac{s_\Delta h_{ij}}{M_\Delta^2}\right)^{-2}$\\ \hline
{\phantom{\Large{l}}}\raisebox{+.2cm}{\phantom{\Large{j}}}
$B_c^-\rightarrow \bar D^0 \pi^+e^-e^-$&$2.21\times 10^{-14}$&$B_c^-\rightarrow \bar D^0 \rho^+e^-e^-$&$4.71\times 10^{-14}$\\
{\phantom{\Large{l}}}\raisebox{+.2cm}{\phantom{\Large{j}}}
$B_c^-\rightarrow \bar D^0 \pi^+\mu^-\mu^-$&$2.36\times 10^{-14}$&$B_c^-\rightarrow \bar D^0 \rho^+\mu^-\mu^-$&$5.05\times 10^{-14}$\\
{\phantom{\Large{l}}}\raisebox{+.2cm}{\phantom{\Large{j}}}
$B_c^-\rightarrow \bar D^0 \pi^+\mu^-e^-$&$4.56\times 10^{-14}$&$B_c^-\rightarrow \bar D^0 \rho^+\mu^-e^-$&$9.73\times 10^{-14}$\\
{\phantom{\Large{l}}}\raisebox{+.2cm}{\phantom{\Large{j}}}
$B_c^-\rightarrow \bar D^0 K^+e^-e^-$&$1.77\times 10^{-15}$&$B_c^-\rightarrow \bar D^0 K^{\ast+}e^-e^-$&$2.67\times 10^{-15}$\\
{\phantom{\Large{l}}}\raisebox{+.2cm}{\phantom{\Large{j}}}
$B_c^-\rightarrow \bar D^0 K^+\mu^-\mu^-$&$1.90\times 10^{-15}$&$B_c^-\rightarrow \bar D^0 K^{\ast+}\mu^-\mu^-$&$2.87\times 10^{-15}$\\
{\phantom{\Large{l}}}\raisebox{+.2cm}{\phantom{\Large{j}}}
$B_c^-\rightarrow \bar D^0 K^+\mu^-e^-$&$3.66\times 10^{-15}$&$B_c^-\rightarrow \bar D^0 K^{\ast+}\mu^-e^-$&$5.52\times 10^{-15}$\\
{\phantom{\Large{l}}}\raisebox{+.2cm}{\phantom{\Large{j}}}
$B_c^-\rightarrow \bar D^0 D^+e^-e^-$&$3.51\times 10^{-13}$&$B_c^-\rightarrow \bar D^0 D^{\ast+}e^-e^-$&$3.38\times 10^{-13}$\\
{\phantom{\Large{l}}}\raisebox{+.2cm}{\phantom{\Large{j}}}
$B_c^-\rightarrow \bar D^0 D^+\mu^-\mu^-$&$3.84\times 10^{-13}$&$B_c^-\rightarrow \bar D^0 D^{\ast+}\mu^-\mu^-$&$3.69\times 10^{-13}$\\
{\phantom{\Large{l}}}\raisebox{+.2cm}{\phantom{\Large{j}}}
$B_c^-\rightarrow \bar D^0 D^+\mu^-e^-$&$7.32\times 10^{-13}$&$B_c^-\rightarrow \bar D^0 D^{\ast+}\mu^-e^-$&$7.03\times 10^{-13}$\\
{\phantom{\Large{l}}}\raisebox{+.2cm}{\phantom{\Large{j}}}
$B_c^-\rightarrow \bar D^0 D_s^+e^-e^-$&$8.23\times 10^{-14}$&$B_c^-\rightarrow \bar D^0 D_s^{\ast+}e^-e^-$&$5.61\times 10^{-14}$\\
{\phantom{\Large{l}}}\raisebox{+.2cm}{\phantom{\Large{j}}}
$B_c^-\rightarrow \bar D^0 D_s^+\mu^-\mu^-$&$9.03\times 10^{-14}$&$B_c^-\rightarrow \bar D^0 D_s^{\ast+}\mu^-\mu^-$&$6.13\times 10^{-14}$\\
{\phantom{\Large{l}}}\raisebox{+.2cm}{\phantom{\Large{j}}}
$B_c^-\rightarrow \bar D^0 D_s^+\mu^-e^-$&$1.72\times 10^{-13}$&$B_c^-\rightarrow \bar D^0 D_s^{\ast+}\mu^-e^-$&$1.17\times 10^{-13}$\\
\hline\hline
\end{tabular*}
\end{table}

\begin{table}
\caption{Branching ratios of $B_c^-$ decays induced by the current $(\bar ub)_{_{V-A}}(\bar q_1q_2)_{_{V-A}}$.}
\vspace{0.2cm}
\setlength{\tabcolsep}{0.01cm}
\centering
\begin{tabular*}{\textwidth}{@{}@{\extracolsep{\fill}}cccc}
\hline\hline
decay channel&$Br\times\left(\frac{s_\Delta h_{ij}}{M_\Delta^2}\right)^{-2}$&decay channel&$Br\times\left(\frac{s_\Delta h_{ij}}{M_\Delta^2}\right)^{-2}$\\ \hline
{\phantom{\Large{l}}}\raisebox{+.2cm}{\phantom{\Large{j}}}
$B_c^-\rightarrow \bar D^{\ast0} \pi^+e^-e^-$&$3.21\times 10^{-16}$&$B_c^-\rightarrow \bar D^{\ast0} \rho^+e^-e^-$&$7.77\times 10^{-16}$\\
{\phantom{\Large{l}}}\raisebox{+.2cm}{\phantom{\Large{j}}}
$B_c^-\rightarrow \bar D^{\ast0} \pi^+\mu^-\mu^-$&$3.43\times 10^{-16}$&$B_c^-\rightarrow \bar D^{\ast0} \rho^+\mu^-\mu^-$&$8.43\times 10^{-16}$\\
{\phantom{\Large{l}}}\raisebox{+.2cm}{\phantom{\Large{j}}}
$B_c^-\rightarrow \bar D^{\ast0} \pi^+\mu^-e^-$&$6.61\times 10^{-16}$&$B_c^-\rightarrow \bar D^{\ast0} \rho^+\mu^-e^-$&$1.62\times 10^{-15}$\\
{\phantom{\Large{l}}}\raisebox{+.2cm}{\phantom{\Large{j}}}
$B_c^-\rightarrow \bar D^{\ast0} K^+e^-e^-$&$4.54\times 10^{-17}$&$B_c^-\rightarrow \bar D^{\ast0} K^{\ast+}e^-e^-$&$6.76\times 10^{-17}$\\
{\phantom{\Large{l}}}\raisebox{+.2cm}{\phantom{\Large{j}}}
$B_c^-\rightarrow \bar D^{\ast0} K^+\mu^-\mu^-$&$4.87\times 10^{-17}$&$B_c^-\rightarrow \bar D^{\ast0} K^{\ast+}\mu^-\mu^-$&$7.33\times 10^{-17}$\\
{\phantom{\Large{l}}}\raisebox{+.2cm}{\phantom{\Large{j}}}
$B_c^-\rightarrow \bar D^{\ast0} K^+\mu^-e^-$&$9.37\times 10^{-17}$&$B_c^-\rightarrow \bar D^{\ast0} K^{\ast+}\mu^-e^-$&$1.41\times 10^{-16}$\\
{\phantom{\Large{l}}}\raisebox{+.2cm}{\phantom{\Large{j}}}
$B_c^-\rightarrow \bar D^{\ast0} D^+e^-e^-$&$2.78\times 10^{-14}$&$B_c^-\rightarrow \bar D^{\ast0} D^{\ast+}e^-e^-$&$5.51\times 10^{-14}$\\
{\phantom{\Large{l}}}\raisebox{+.2cm}{\phantom{\Large{j}}}
$B_c^-\rightarrow \bar D^{\ast0} D^+\mu^-\mu^-$&$3.05\times 10^{-14}$&$B_c^-\rightarrow \bar D^{\ast0} D^{\ast+}\mu^-\mu^-$&$6.08\times 10^{-14}$\\
{\phantom{\Large{l}}}\raisebox{+.2cm}{\phantom{\Large{j}}}
$B_c^-\rightarrow \bar D^{\ast0} D^+\mu^-e^-$&$5.79\times 10^{-14}$&$B_c^-\rightarrow \bar D^{\ast0} D^{\ast+}\mu^-e^-$&$1.15\times 10^{-13}$\\
{\phantom{\Large{l}}}\raisebox{+.2cm}{\phantom{\Large{j}}}
$B_c^-\rightarrow \bar D^{\ast0} D_s^+e^-e^-$&$6.65\times 10^{-15}$&$B_c^-\rightarrow \bar D^{\ast0} D_s^{\ast+}e^-e^-$&$1.01\times 10^{-14}$\\
{\phantom{\Large{l}}}\raisebox{+.2cm}{\phantom{\Large{j}}}
$B_c^-\rightarrow \bar D^{\ast0} D_s^+\mu^-\mu^-$&$7.32\times 10^{-15}$&$B_c^-\rightarrow \bar D^{\ast0} D_s^{\ast+}\mu^-\mu^-$&$1.12\times 10^{-14}$\\
{\phantom{\Large{l}}}\raisebox{+.2cm}{\phantom{\Large{j}}}
$B_c^-\rightarrow \bar D^{\ast0} D_s^+\mu^-e^-$&$1.39\times 10^{-14}$&$B_c^-\rightarrow \bar D^{\ast0} D_s^{\ast+}\mu^-e^-$&$2.12\times 10^{-14}$\\
\hline\hline
\end{tabular*}
\end{table}

\begin{table}
\caption{Branching ratios of $B_c^-$ decays induced by the current $(\bar q_1c)_{_{V-A}}(\bar q_2q_3)_{_{V-A}}$.}
\vspace{0.2cm}
\setlength{\tabcolsep}{0.01cm}
\centering
\begin{tabular*}{\textwidth}{@{}@{\extracolsep{\fill}}cccc}
\hline\hline
decay channel& $Br\times\left(\frac{s_\Delta h_{ij}}{M_\Delta^2}\right)^{-2}$&decay channel& $Br\times\left(\frac{s_\Delta h_{ij}}{M_\Delta^2}\right)^{-2}$\\ \hline
{\phantom{\Large{l}}}\raisebox{+.2cm}{\phantom{\Large{j}}}
$B_c^-\rightarrow \bar B^0 \pi^+ e^-e^-$&$2.90\times 10^{-13}$&$B_c^-\rightarrow \bar B^0 \rho^+ e^-e^-$&$7.74\times 10^{-15}$\\
{\phantom{\Large{l}}}\raisebox{+.2cm}{\phantom{\Large{j}}}
$B_c^-\rightarrow \bar B^0 \pi^+ \mu^-\mu^-$&$2.84\times 10^{-13}$&$B_c^-\rightarrow \bar B^0 \rho^+ \mu^-\mu^-$&$1.60\times 10^{-17}$\\
{\phantom{\Large{l}}}\raisebox{+.2cm}{\phantom{\Large{j}}}
$B_c^-\rightarrow \bar B^0 \pi^+ \mu^-e^-$&$5.44\times 10^{-13}$&$B_c^-\rightarrow \bar B^0 \rho^+ \mu^-e^-$&$3.56\times 10^{-15}$\\
{\phantom{\Large{l}}}\raisebox{+.2cm}{\phantom{\Large{j}}}
$B_c^-\rightarrow \bar B^0 K^+ e^-e^-$&$2.24\times 10^{-13}$&$B_c^-\rightarrow \bar B^0 K^{\ast+} e^-e^-$&$3.28\times 10^{-16}$\\
{\phantom{\Large{l}}}\raisebox{+.2cm}{\phantom{\Large{j}}}
$B_c^-\rightarrow \bar B^0 K^+ \mu^-\mu^-$&$2.04\times 10^{-13}$&$B_c^-\rightarrow \bar B^0 K^{\ast+} \mu^-\mu^-$&$-$\\
{\phantom{\Large{l}}}\raisebox{+.2cm}{\phantom{\Large{j}}}
$B_c^-\rightarrow \bar B^0 K^+ \mu^-e^-$&$3.88\times 10^{-13}$&$B_c^-\rightarrow \bar B^0 K^{\ast+} \mu^-e^-$&$-$\\
{\phantom{\Large{l}}}\raisebox{+.2cm}{\phantom{\Large{j}}}
$B_c^-\rightarrow \bar B_s^0 \pi^+ e^-e^-$&$1.41\times 10^{-12}$&$B_c^-\rightarrow \bar B_s^0 \rho+ e^-e^-$&$6.63\times 10^{-15}$\\
{\phantom{\Large{l}}}\raisebox{+.2cm}{\phantom{\Large{j}}}
$B_c^-\rightarrow \bar B_s^0 \pi^+ \mu^-\mu^-$&$1.31\times 10^{-12}$&$B_c^-\rightarrow \bar B_s^0 \rho^+ \mu^-\mu^-$&$-$\\
{\phantom{\Large{l}}}\raisebox{+.2cm}{\phantom{\Large{j}}}
$B_c^-\rightarrow \bar B_s^0 \pi^+ \mu^-e^-$&$2.55\times 10^{-12}$&$B_c^-\rightarrow \bar B_s^0 \rho^+ \mu^-e^-$&$9.89\times 10^{-17}$\\
{\phantom{\Large{l}}}\raisebox{+.2cm}{\phantom{\Large{j}}}
$B_c^-\rightarrow \bar B_s^0 K^+ e^-e^-$&$1.02\times 10^{-13}$&$B_c^-\rightarrow \bar B_s^0 K^{\ast}+ e^-e^-$&$8.26\times 10^{-20}$\\
{\phantom{\Large{l}}}\raisebox{+.2cm}{\phantom{\Large{j}}}
$B_c^-\rightarrow \bar B_s^0 K^+ \mu^-\mu^-$&$7.80\times 10^{-14}$&$B_c^-\rightarrow \bar B_s^0 K^{\ast+} \mu^-\mu^-$&$-$\\
{\phantom{\Large{l}}}\raisebox{+.2cm}{\phantom{\Large{j}}}
$B_c^-\rightarrow \bar B_s^0 K^+ \mu^-e^-$&$1.58\times 10^{-13}$&$B_c^-\rightarrow \bar B_s^0 K^{\ast+} \mu^-e^-$&$-$\\
\hline\hline
\end{tabular*}
\end{table}

\begin{table}
\caption{Branching ratios of $B_c^-$ decays induced by the current $(\bar q_1c)_{_{V-A}}(\bar q_2q_3)_{_{V-A}}$.}
\vspace{0.2cm}
\setlength{\tabcolsep}{0.01cm}
\centering
\begin{tabular*}{\textwidth}{@{}@{\extracolsep{\fill}}cccc}
\hline\hline
decay channel& $Br\times\left(\frac{s_\Delta h_{ij}}{M_\Delta^2}\right)^{-2}$&decay channel& $Br\times\left(\frac{s_\Delta h_{ij}}{M_\Delta^2}\right)^{-2}$\\ \hline
{\phantom{\Large{l}}}\raisebox{+.2cm}{\phantom{\Large{j}}}
$B_c^-\rightarrow \bar B^{\ast0} \pi^+ e^-e^-$&$1.03\times 10^{-14}$&$B_c^-\rightarrow \bar B_s^{\ast0} \pi^+ e^-e^-$&$1.20\times 10^{-13}$\\
{\phantom{\Large{l}}}\raisebox{+.2cm}{\phantom{\Large{j}}}
$B_c^-\rightarrow \bar B^{\ast0} \pi^+ \mu^-\mu^-$&$1.11\times 10^{-14}$&$B_c^-\rightarrow \bar B_s^{\ast0} \pi^+ \mu^-\mu^-$&$1.28\times 10^{-13}$\\
{\phantom{\Large{l}}}\raisebox{+.2cm}{\phantom{\Large{j}}}
$B_c^-\rightarrow \bar B^{\ast0} \pi^+ \mu^-e^-$&$2.04\times 10^{-14}$&$B_c^-\rightarrow \bar B_s^{\ast0} \pi^+ \mu^-e^-$&$2.37\times 10^{-13}$\\
{\phantom{\Large{l}}}\raisebox{+.2cm}{\phantom{\Large{j}}}
$B_c^-\rightarrow \bar B^{\ast0} K^+ e^-e^-$&$3.18\times 10^{-14}$&$B_c^-\rightarrow \bar B_s^{\ast0} K^+ e^-e^-$&$4.31\times 10^{-14}$\\
{\phantom{\Large{l}}}\raisebox{+.2cm}{\phantom{\Large{j}}}
$B_c^-\rightarrow \bar B^{\ast0} K^+ \mu^-\mu^-$&$2.91\times 10^{-14}$&$B_c^-\rightarrow \bar B_s^{\ast0} K^+ \mu^-\mu^-$&$3.92\times 10^{-14}$\\
{\phantom{\Large{l}}}\raisebox{+.2cm}{\phantom{\Large{j}}}
$B_c^-\rightarrow \bar B^{\ast0} K^+ \mu^-e^-$&$5.48\times 10^{-14}$&$B_c^-\rightarrow \bar B_s^{\ast0} K^+ \mu^-e^-$&$7.40\times 10^{-14}$\\
\hline\hline
\end{tabular*}
\end{table}

\section{Conclusions}

We have studied the doubly-charged Higgs boson induced lepton number violation processes of $B_c$ meson. Both the three-body decay channels and four-body decay channels are considered. For the former, the largest value of $Br\times\left(\frac{s_\Delta h_{ij}}{M_\Delta^2}\right)^{-2}$ is of the order of $10^{-7}$, which comes from the $D_s^+l^-l^-$ channel. For the later ones, $Br\times\left(\frac{s_\Delta h_{ij}}{M_\Delta^2}\right)^{-2}$ is of the order of $10^{-12}\sim 10^{-20}$. The largest value comes from the $\bar B_s^{0}\pi^+l^-l^-$ channels. But they are still three orders smaller than the smallest value of three-body decay channels. The branching ratios of these channels are much smaller than the experimental precision, which makes them no possible to be achieved in the current experiments. However, our work could be a helpful supplement for the studies of neutrinoless double beta decay processes of $B_c$ meson.

\section*{ACKNOWLEDGEMENTS}
This work was supported in part by the National Natural Science
Foundation of China (NSFC) under Grant No.~11405037, No.~11575048, No.~11505039, and No.~11405004, and in part by PIRS of HIT No. B201506.
\section*{Appendix}

The hadronic transition amplitude can be written as~\cite{chang06}
\begin{equation}
\begin{aligned}
\langle h_1^0(p_1)|\bar q_1\gamma^\mu(1-\gamma_5)b|B_c^-(p)\rangle=\int\frac{d{\vec{q}}}{(2\pi)^{3}}\textrm{Tr}\left[\frac{\slashed p}{M}\overline{\varphi_{p_1}^{++}}({\vec{q}_1})\gamma_{\mu}(1-\gamma_{5})\varphi_{p}^{++}({\vec{q}})\right],
\end{aligned}
\end{equation}
where $\vec q$ and $\vec q_1$ are the relative momenta of $B_c^-$ and $h_1^0$ mesons, repectively.
$\varphi_{p}^{++}({\vec{q}})$ and $\varphi_{p_1}^{++}({\vec{q_1}})$ are the positive energy parts of the wave functions of the initial and final heavy mesons, respectively, which have the following forms~\cite{wang17}
\begin{equation}
\begin{aligned}
\varphi^{++}_{0^-}(q_\perp)&=\left[A_1(q_\perp)+\frac{\slashed{P}}{M}A_2(q_\perp)
+\frac{\slashed{q}_\perp}{M}A_3(q_\perp)+\frac{\slashed{P}\slashed{q}_\perp}{M^2}A_4(q_\perp)\right]\gamma_5,
\end{aligned}
\end{equation}
where the coefficients are
\begin{equation}
\begin{aligned}
&A_1 = \frac{M}{2}\left[\frac{\omega_1+\omega_2}{m_1+m_2}f_1+f_2\right],\\
&A_2=\frac{M}{2}\left[f_1+\frac{m_1+m_2}{\omega_1+\omega_2}f_2\right],\\
&A_3=-\frac{M(\omega_1-\omega_2)}{m_1\omega_2+m_2\omega_1}A_1,\\
&A_4=-\frac{M(m_1+m_2)}{m_1\omega_2+m_2\omega_1}A_1.
\end{aligned}
\end{equation}
In the above equation, $m_1$ and $m_2$ are respectively the masses of quark and antiquark inside the meson. $\omega_i$ is defined as $\sqrt{m_i^2+\vec q^2}$. $f_1$ and $f_2$ are functions of $\vec q^2$.

For the $1^-$ state, the positive energy part of the wave function has the form~\cite{wang17}
\begin{equation}
\begin{aligned}
\varphi^{++}_{1^-}(q_\perp)&=(q_\perp\cdot\epsilon)\left[B_1(q_\perp)+\frac{\slashed{P}}{M}B_2(q_\perp)
+\frac{\slashed{q}_\perp}{M}B_3(q_\perp)+\frac{\slashed{P}\slashed{q}_\perp}{M^2}B_4(q_\perp)\right]\\
&~~~~~+ M\slashed\epsilon\left[B_5(q_\perp)+\frac{\slashed{P}}{M}B_6(q_\perp)
+\frac{\slashed{q}_\perp}{M}B_7(q_\perp)+\frac{\slashed{P}\slashed{q}_\perp}{M^2}B_8(q_\perp)\right],
\end{aligned}
\end{equation}
where the coefficients are
\begin{equation}
\begin{aligned}
&B_1 = \frac{1}{2M(m_1\omega_2 + m_2\omega_1)}\left[(\omega_1+\omega_2) q_\perp^2f_3 + (m_1 + m_2) q_\perp^2f_4 + 2M^2\omega_2f_5 -2M^2m_2f_6\right],\\
&B_2 = \frac{1}{2M(m_1\omega_2 + m_2\omega_1)}\left[(m_1-m_2) q_\perp^2f_3 + (\omega_1 - \omega_2)q_\perp^2f_4 - 2M^2m_2f_5 +2M^2\omega_2f_6\right],\\
&B_3 = \frac{1}{2}\left[f_3 + \frac{m_1 + m_2}{\omega_1+\omega_2}f_4 -\frac{2M^2}{m_1\omega_2  + m_2\omega_1}f_6\right],\\
&B_4 = \frac{1}{2}\left[\frac{\omega_1+\omega_2}{m_1+m_2}f_3 + f_4 -\frac{2M^2}{m_1\omega_2+m_2\omega_1}f_5\right],\\
&A_5 = \frac{1}{2}\left[f_5 -\frac{\omega_1+\omega_2}{m_1+m_2}f_6\right],~~~~~~~~~~~~A_6 = \frac{1}{2}\left[-\frac{m_1+m_2}{\omega_1+\omega_2}f_5 +f_6\right],\\
&B_7 = \frac{M}{2}\frac{\omega_1-\omega_2}{m_1\omega_2+m_2\omega_1}\left[f_5 - \frac{\omega_1+\omega_2}{m_1+m_2}f_6\right],\\
&B_8 = \frac{M}{2}\frac{m_1+m_2}{m_1\omega_2 + m_2\omega_1}\left[-f_5 +\frac{\omega_1+\omega_2}{m_1+m_2}f_6\right].
\end{aligned}
\end{equation}


\end{document}